\documentclass[aps,prl,reprint]{revtex4-1}
\usepackage[separate-uncertainty=true]{siunitx}
\usepackage{amsmath}
\usepackage{graphicx}

\renewcommand{\vec}[1]{\mathbf{#1}}
\newcommand{\abs}[1]{\left\vert #1 \right\vert}

\begin{document}

\title{Interpreting Holographic Molecular Binding Assays 
with Effective Medium Theory}

\author{Lauren E. Altman}
\author{David G. Grier}

\affiliation{Department of Physics and Center for Soft Matter Research, New York University, New York, NY 10003, USA}

\begin{abstract}
Holographic molecular binding assays use holographic video
microscopy to directly detect molecules binding 
to the surfaces of micrometer-scale colloidal beads by monitoring 
associated changes in the beads' light-scattering properties.
Holograms of individual spheres are analyzed
by fitting to a generative model
based on the Lorenz-Mie theory of light scattering.
Each fit yields an estimate of a probe 
bead's diameter and refractive index with
sufficient precision to watch the beads
grow as molecules bind.
Rather than modeling the molecular-scale coating, however,
these fits use effective medium theory, treating the
coated sphere as if it were homogeneous.
This effective-sphere analysis is rapid and numerically robust
and so is useful for practical implementations 
of label-free immunoassays.
Here, we assess how effective-sphere properties
reflect the properties of molecular-scale coatings
by modeling coated spheres
with the discrete-dipole approximation and analyzing
their holograms with the effective-sphere model.
\end{abstract}

\maketitle

\section{Introduction}

\begin{figure*}
    \centering
    \includegraphics[width=0.75\textwidth]{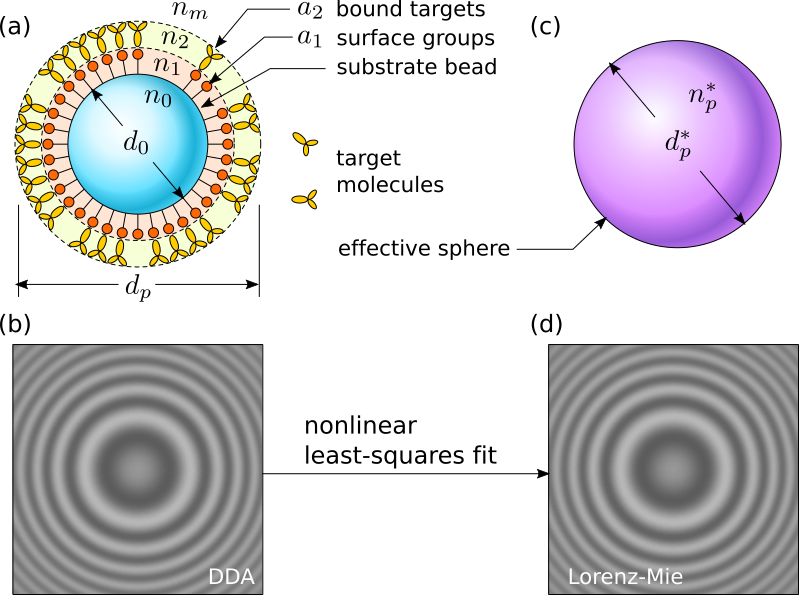}
    \caption{(a) Geometry of a bead-based molecular binding
    assay. A substrate bead with diameter $d_0$ 
    and refractive index $n_0$ is coated with a layer of
    functional groups that bind target molecules from the surrounding solution. The surface groups are modeled as a dielectric
    layer of thickness $a_1$ and refractive index $n_1$.
    The coating of bound target molecules is modeled as a
    second layer of thickness $a_2$ and refractive index $n_2$.
    The bead is dispersed in a medium of refractive index $n_m$.
    The bounding sphere has diameter $d_p = d_0 + 2(a_1 + a_2)$.
    (b) Synthetic hologram of a coated bead computed 
    with the DDA. Parameters are chosen to mimic a
    typical molecular binding assay based on a probe
    bead with a polystyrene core:
    $d_0 = \SI{1}{\um}$, $n_0 = \num{1.60}$, 
    and molecular overlayers with
    $a_1 = a_2 = \SI{10}{\nm}$, $n_1 = n_2 = \num{1.55}$.
    (c) The effective-sphere model treats the coated sphere
    as if it were a homogeneous dielectric sphere of effective
    diameter $d_p^\ast$ and refractive index $n_p^\ast$. 
    (d) Fitting the coated-sphere's hologram from (b) with
    the Lorenz-Mie theory for a homogeneous sphere yields an indistinguishable hologram for 
    $d_p^\ast = \SI{1.03}{\um}$ and $n_p^\ast = \num{1.594}$.}
    \label{fig:geometry}
\end{figure*}

Holographic particle characterization 
works by recording the in-line hologram
of a colloidal particle \cite{sheng_digital_2006}
and then analyzing it pixel-by-pixel \cite{lee_characterizing_2007} with a generative model based on 
the Lorenz-Mie theory of light scattering \cite{bohren_absorption_1983,mishchenko_scattering_2002,gouesbet_generalized_2011}.
This analysis yields the particle's diameter with
nanometer-scale precision and its refractive index to
within a part per thousand 
\cite{lee_characterizing_2007,krishnatreya_measuring_2014},
which should be fine enough to detect
nanometer-scale molecules binding to the surfaces
of micrometer-scale colloidal beads 
\cite{cheong_flow_2009,zagzag_holographic_2020}.
The resulting label-free assay for molecular binding,
shown schematically in Fig.~\ref{fig:geometry}(a),
has immediate applications for medical diagnostics,
a proof-of-concept immunoassay based on holographic
particle characterization having recently been reported
\cite{zagzag_holographic_2020}.
Holographic molecular binding assays
currently are being developed into
diagnostic and serological tests
for infection by SARS-CoV-2, the virus responsible for COVID-19.

Practical implementations of holographic molecular binding assays
rely on the effective-sphere model
\cite{cheong_flow_2009,hannel_holographic_2015,wang_holographic_2016,wang_holographic_2016-1,fung_computational_2019,odete_role_2020},
which treats the scatterer as a homogeneous isotropic sphere
regardless of its actual composition and microstructure.
Values obtained for the particle's 
diameter and refractive index
then are interpreted to be characteristics of an effective sphere
whose properties represent averages over the particle's
inhomogeneities using Maxwell Garnett effective-medium
theory \cite{markel_introduction_2016}.

The effective-sphere 
model has been validated through
studies on porous spheres \cite{cheong_holographic_2011,odete_role_2020}
and fractal colloidal clusters \cite{wang_holographic_2016,wang_holographic_2016-1,cheong_holographic_2017,fung_computational_2019}.
In both cases, the two phases that make up
the particle are distributed uniformly,
which is consistent with
the assumptions underlying
Maxwell Garnett theory.
The heterogeneity in bead-based
molecular binding assays, by contrast, is
restricted to thin surface layers.
The present study assesses how properties
of coated spheres
estimated with the effective-sphere model
reflect the presence and composition of
the surface layers with the goal of
guiding the development of fast and
effective holographic molecular binding assays.

We appraise the effective-sphere
analysis of coated beads
by using it to analyze synthetic holograms 
computed with the discrete-dipole approximation
(DDA) \cite{draine_discrete-dipole_1994}.
Direct comparisons between ground truth values
and fits to the effective-sphere model demonstrate
that effective-sphere analysis usefully
characterizes coated spheres, reliably detecting
the presence of coatings and offering insights
into their properties.
The results of this study are consistent with
trends in bead diameter and refractive index
reported in experimental demonstrations of
holographic molecular binding assays.
This positive outcome furthermore means that 
molecular binding assays based on holographic particle characterization can benefit from
the speed and robustness
of effective-sphere analysis.

\subsection{Holographic particle characterization}
\label{sec:holographicparticlecharacterization}

The holograms used for holographic particle 
characterization are recorded by illuminating
a colloidal dispersion with a
collimated laser
beam \cite{sheng_digital_2006,lee_characterizing_2007}.
Light scattered by a colloidal particle
interferes with the remainder of the beam in
the focal plane of a microscope that
magnifies the interference pattern
and relays it to a video camera.
Each magnified intensity pattern recorded
by the camera is a hologram of the particles
in the observation volume and encodes information
on their three-dimensional
positions, as well as their sizes, shapes
and compositions.

Holographic particle characterization extracts
information from recorded holograms by fitting
to a generative model
for the image-formation process \cite{lee_characterizing_2007}.
A standard implementation models
the incident beam as a unit-amplitude
monochromatic plane wave at frequency $\omega$,
\begin{equation}
\label{eq:incidentbeam}
    \vec{E}_0(\vec{r}, t)
    =
    e^{i k z} e^{-i \omega t} \, \hat{x},
\end{equation}
that is linearly polarized along $\hat{x}$
and propagates along $\hat{z}$.
The wavenumber, $k = 2 \pi n_m/\lambda$,
depends on the laser's vacuum wavelength, $\lambda$,
and the refractive index of the medium,
$n_m$.
This beam illuminates a particle
located at $\vec{r}_p$ relative to the
center of the microscope's focal plane.
The time-averaged
intensity pattern recorded by the camera
therefore may be modeled as
\begin{equation}
\label{eq:b}
    b(\vec{r})
    =
    \abs{
    \hat{x} + 
    e^{-i k z_p} \,
    \vec{f}_s(k(\vec{r} - \vec{r}_p))
    }^2,
\end{equation}
where $\vec{f}_s(k\vec{r})$ is the Lorenz-Mie
scattering function for the particle \cite{bohren_absorption_1983,mishchenko_scattering_2002,gouesbet_generalized_2011}.
In practice, an experimentally recorded hologram
is corrected for the dark count of the camera
and normalized by the intensity distribution
of the illumination
to facilitate comparison with Eq.~\eqref{eq:b}.

The scattering function for a homogeneous sphere
is parametrized by
the sphere's diameter, $d_p$, and refractive
index $n_p$.
Fitting a single-particle hologram to
Eq.~\eqref{eq:b} involves optimizing these two parameters
plus the particle's three-dimensional position,
$\vec{r}_p$.
Published implementations
\cite{wiscombe_improved_1980,yang_improved_2003,pena_scattering_2009,krishnatreya_fast_2014,altman_catch_2020}
can localize and characterize a sphere
in a typical video image and
in under 50 milliseconds
on a desktop workstation.

Our numerical studies are performed with parameters
appropriate for the commercial implementation of
holographic particle characterization (xSight, Spheryx, Inc.) .
This platform currently is being used
to develop holographic antibody binding assays of the kind depicted in Fig.~\ref{fig:geometry}.
It operates at a vacuum wavelength
of $\lambda = \SI{445}{\nm}$ and has
an effective system magnification of
\SI{120}{\nm\per pixel}.
We furthermore assume that the medium has the
refractive index of water at the imaging
wavelength, $n_m = \num{1.340}$.
No other calibration constants are required.

Validation experiments on colloidal size standards
demonstrate that holographic particle characterization
measurements, including those performed with xSight,
can resolve the diameter of a micrometer-scale
sphere with a precision of \SI{5}{\nm} \cite{lee_holographic_2007}. 
Measurements on emulsion droplets demonstrate
precision and reproducibility in the refractive
index of \num{0.001} \cite{shpaisman_holographic_2012}.
The former should suffice to
detect the formation of a molecular coating through
the associated change in the bead's diameter
\cite{cheong_flow_2009,zagzag_holographic_2020},
while the latter is useful for distinguishing different
types of beads on the basis of their
composition \cite{yevick_machine-learning_2014}.

In principle, the hologram of a coated sphere
could be analyzed by suitably generalizing
the scattering function, $\vec{f}_s(k\vec{r})$,
to account for the thicknesses and
refractive indexes of its coatings
\cite{yang_improved_2003,pena_scattering_2009}.
Introducing these additional adjustable
parameters, however, reduces the likelihood
that the fits will converge successfully
and increases the measurement's sensitivity
to noise and uncorrected interference artifacts
in the recorded images.
The extracted values for the parameters, moreover, 
still would reflect effective-medium 
characterizations of
molecular overlayers that could
be patchy or incomplete.
We therefore seek to understand how
coatings influence effective-sphere parameters,
which can be obtained rapidly and reliably.

\section{Testing effective-sphere analysis
of coated spheres with the
discrete-dipole approximation}

To assess how effective-sphere measurements reflect
the actual properties of a coated sphere, we 
numerically compute the ideal hologram of a coated sphere
using the discrete-dipole approximation, and then
analyze the hologram using Lorenz-Mie theory for a
homogeneous sphere.
The discrete-dipole approximation (DDA) 
\cite{draine_discrete-dipole_1994,yurkin_discrete_2007,yurkin_discrete-dipole-approximation_2011}
treats a scatterer as an ensemble of point-like dipoles.
Each elementary dipole
scatters the incident plane wave, redirecting
a portion to the imaging plane.
Some of the scattered light reaches neighboring
dipoles, which scatter it a second time.
Some of that twice-scattered light also reaches
the imaging plane and contributes to the computed image.
The first-order DDA truncates the hierarchy after
the first neighbor-scattering contribution
both to reduce computation time and also to
maintain numerical stability.

Our implementation
uses the \texttt{holopy} interface
\cite{barkley_holographic_2019}
to the ADDA library \cite{yurkin_discrete-dipole-approximation_2011}.
We model a coated sphere by specifying the
properties of the individual dipoles in a discrete
three-dimensional lattice with 
$10 n_0/n_m$ dipoles per wavelength, $\lambda$,
along each axis.
Dipoles located within the substrate sphere are assigned
refractive indexes $n_0$.
Those within coatings are assigned $n_1$ or
$n_2$, as indicated in Fig.~\ref{fig:geometry}(a).
The field scattered by this system of dipoles
replaces $\vec{f}_s(k\vec{r})$ in Eq.~\eqref{eq:b}
in computing the ideal hologram.
The example in Fig.~\ref{fig:geometry}(b)
was computed in this way.

Our effective-sphere analysis is performed with the
\texttt{pylorenzmie} software suite that automates
fits to Eq.~\eqref{eq:b}.
The hologram in Fig.~\ref{fig:geometry}(d) depicts
the effective-sphere model's best fit the DDA hologram
in Fig.~\ref{fig:geometry}(b).
Whereas the DDA hologram requires 
\SI{1}{\minute}
to compute, the equivalent effective-sphere
hologram can be computed in under \SI{5}{\ms}
on the same computer hardware.
Given that a single fit to an experimental hologram
can require dozens of realizations, the speed
of the effective-sphere model offers clear advantages
provided that its results can be interpreted
productively.

Using \texttt{pylorenzmie} 
to analyze holograms recorded
by xSight yields characterization results
consistent with those reported by the
instrument's own analytical software \cite{altman_catch_2020}.
We are confident, therefore, that results of our
numerical experiments reflect the performance
of the effective-sphere model for
real-world measurements.

\subsection{Validating DDA and effective-sphere implementations}
\label{sec:validation}

\begin{figure*}
    \centering
    \includegraphics[width = \textwidth]{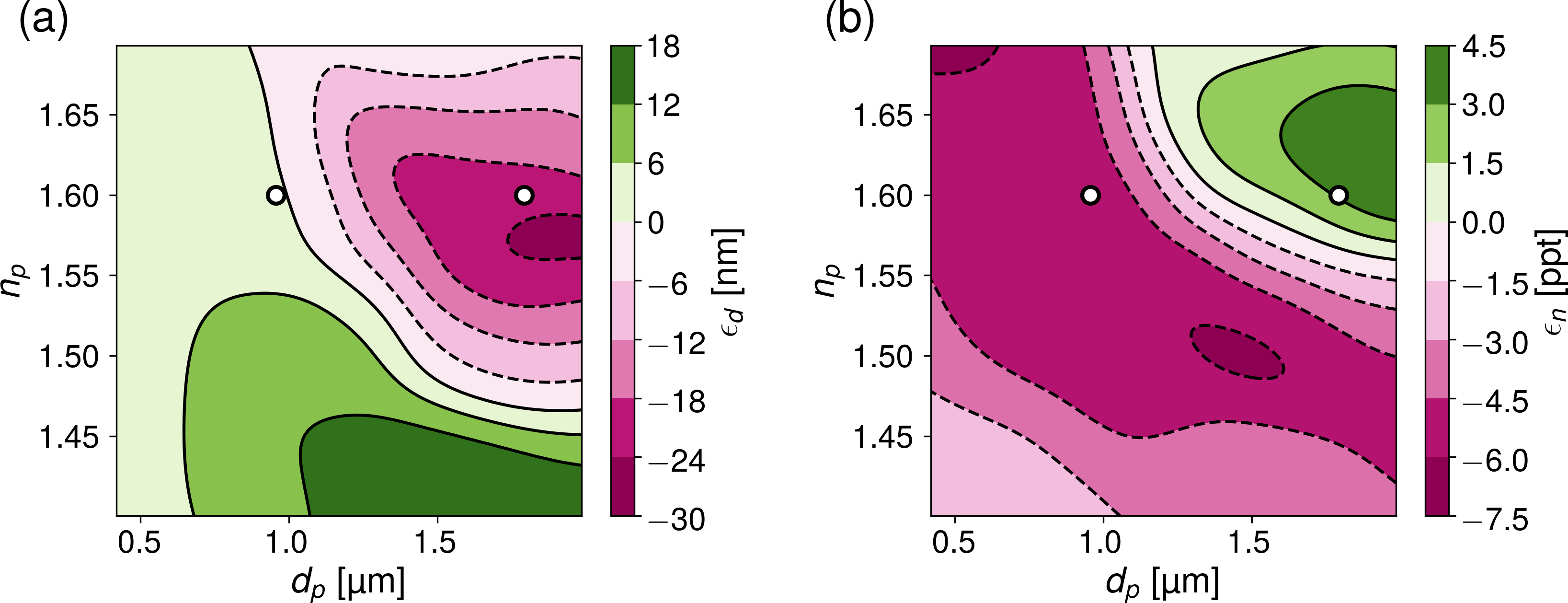}
    \caption{Holographic images of homogeneous spheres computed with DDA are then fitted to Lorenz-Mie theory. Their deviations from the inputted values are $\epsilon_d(d_p, n_p) = d_p^\ast - d_p$ (a), and $\epsilon_n(d_p, n_p) = n_p^\ast - n_p$ (b). Errors can be positive or negative for both parameters, but in this region of parameter space, errors in $n_p$ are predominantly negative. 
    Discrete points indicate the properties of polystyrene
    substrate beads used in published molecular binding
    assays \cite{cheong_flow_2009,zagzag_holographic_2020}
    with diameters  $d_p = \SI{1}{\um}$ and
    $d_p = \SI{1.8}{\um}$.
    The smaller of these lies in a region of parameter space where agreement between DDA and Lorenz-Mie formulations
    is particularly good with $\epsilon_d < \SI{5}{\nm}$.}
    \label{fig:validation}
\end{figure*}

We validate the combination of DDA hologram synthesis
and Lorenz-Mie analysis by performing numerical experiments
on homogeneous spheres
for which Eq.~\eqref{eq:b} should be exact.
In this case,
fitted values for the diameter, $d_p^\ast$, and 
the refractive index, $n_p^\ast$,
should agree with the ground-truth 
values, $d_p$ and $n_p$, used as inputs.
Figure~\ref{fig:validation}(a) presents a map of the
errors in the estimated diameter,
$\epsilon_d(d_p, n_p) = d_p^\ast - d_p$,
as a function of the ground-truth inputs.
Figure~\ref{fig:validation}(b) shows corresponding
results for errors in the refractive index,
$\epsilon_n(d_p, n_p) = n_p^\ast - n_p$.
The DDA and Lorenz-Mie formulations agree on
the spheres' diameters to within
the \SI{+-5}{\nm} precision of holographic
characterization measurements over more than half
of the selected parameter range, the agreement being
better than \SI{1}{\nm} for smaller spheres.
Errors in refractive index generally are
smaller than \SI{+-5}{ppt} across the entire
domain, which also is reasonable for our application.
Errors in both diameter and refractive index
are consistent with previous reports of
the performance of the DDA
\cite{yurkin_discrete_2007,yurkin_discrete-dipole-approximation_2011}.

The discrete data points in Fig.~\ref{fig:validation}(a) and 
Fig.~\ref{fig:validation}(b) represent the properties
of the micrometer-diameter polystyrene spheres that were
used for reported molecular binding assays
\cite{cheong_flow_2009,zagzag_holographic_2020}.
The computational methods' errors in size and refractive index
are comparable to instrumental uncertainties for the
smaller of these particles, but exceed instrumental
uncertainties for the larger.
We focus our numerical study, therefore, on the characteristics
of the smaller commercial probe beads,
both for their immediate practical
application to medical testing
and also to assess the effective-sphere model
under conditions where our techniques are most reliable.

\subsection{Single Coatings}
\label{sec:singlecoat}

We next examine the effect of adding a single 
coating of a homogeneous material onto a uniform substrate
sphere.
This is a model for hologram formation by 
the probe beads used for
holographic molecular binding assays.
To facilitate comparison with recent experimental studies
\cite{cheong_flow_2009,zagzag_holographic_2020},
we focus on the particular case of micrometer-diameter
polystyrene spheres with $d_0 = \SI{1}{\um}$ and $n_0 = \num{1.60}$.
The coating thickness, $a_1$, and refractive index, $n_1$, are 
selected at random from the range 
$a_1 \in [\SI{5}{\nm}, \SI{20}{\nm}]$
and $n_1 \in [\num{1.4}, \num{1.7}]$.
For each set of parameters, we use DDA to compute the coated
sphere's hologram and then fit to the effective-sphere model
for $d_p^\ast$ and $n_p^\ast$.

In the special case that the coating has the same refractive
index as the substrate, $n_1 = n_0$, adding a coating is
equivalent to increasing the diameter of the bead:
$d_p^\ast = d_0 + 2 a_1$.
Alternatively, setting $n_1 = n_m$ is equivalent to not
adding a coating, which should yield $d_p^\ast = d_0$.

\begin{figure*}
    \centering
    \includegraphics[width=0.7\textwidth]{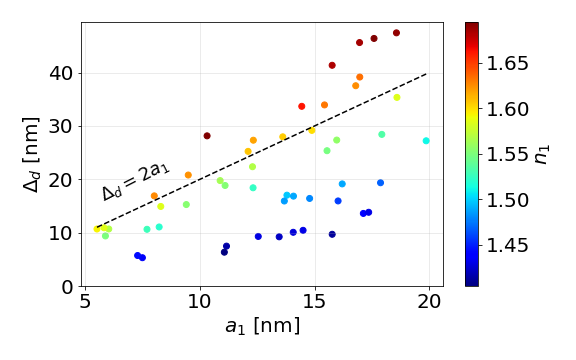}
    \caption{Influence of coating thickness, $a_1$, 
    on effective sphere diameter, $d_p^\ast$, depends on the refractive index of the
    coating, $n_1$, relative to that of the substrate sphere, $n_0$.
    Data points are computed for a sphere with $d_0 = \SI{1}{\um}$
    and $n_0 = \num{1.60}$. The observed increase,
    $\Delta_d(a_1) = d_p^\ast - d_0$, increases monotonically with
    $n_1$ and agrees with the geometric size, $\Delta_d(a_1) = 2 a_1$.
    when $n_1 = n_0$, as indicated by the dashed line.}
    \label{fig:single_coating_size}
\end{figure*}

\begin{figure*}
    \centering
    \includegraphics[width=\textwidth]{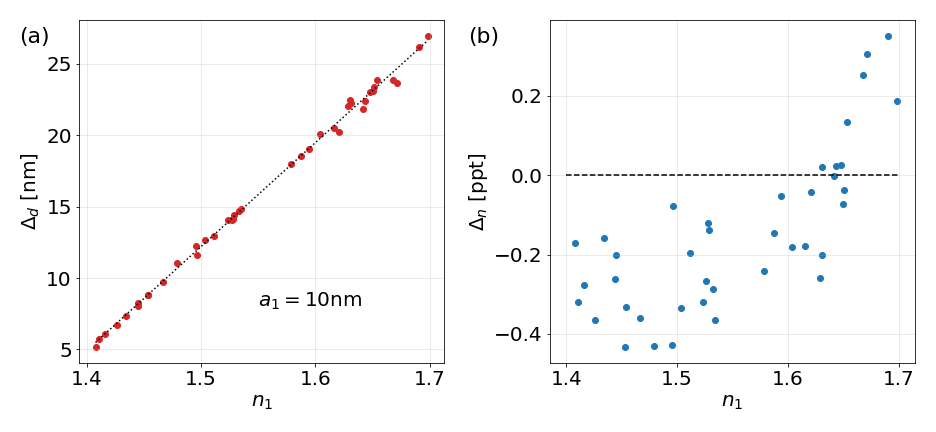}
    \caption{A \SI{10}{\nm} coating with variable refractive index $n_1$ is applied to a $d_0 = \SI{1}{\um}$ polystyrene bead and then is fitted to the effective-sphere model. (a) Shift in effective diameter $\Delta_d$ is linear with coating index $n_1$, with $n_1 = n_p = \num{1.6}$ corresponding to a shift of \SI{10}{nm}. A regression line (dotted) is fitted to the data. to the data. (b) Shift in effective refractive index, $\Delta_n$, is positive for $n_1 > n_p$ and negative for $n_1 < n_p$. The horizontal dashed line indicates
    the baseline, $\Delta_n = 0$.}
    \label{fig:single_coating_index}
\end{figure*}

The results in Fig.~\ref{fig:single_coating_size} confirm
that adding a molecular-scale coating increases the apparent
diameter of the bead, $\Delta_d = d_p^\ast - d_0 > 0$,
provided the coating's refractive index is greater than
that of the medium.
As expected, the coating's influence on the effective diameter depends
on its refractive index relative to that of the substrate,
with low-index coatings increasing $d_p^\ast$ by less than their
thickness and high-index coatings increasing $d_p^\ast$ by more.
Indeed, Fig.~\ref{fig:single_coating_index}(a) 
shows
that the apparent bead diameter increases
nearly linearly with $n_1$ for a fixed layer
thickness, $a_1 = \SI{10}{\nm}$.
Extrapolating to $n_1 = n_m$ yields $\Delta_d = 0$,
as expected.
Similarly, setting $n_1 = n_0$ yields
$\Delta_d = 2 a_1$.
From this, we obtain a phenomenological
relationship between the effective
diameter and the properties
of the coating:
\begin{equation}
    d_p^\ast 
    = 
    d_0 + 2 a_1 \, \frac{n_1 - n_m}{n_0 - n_m}.
\end{equation}

Figure~\ref{fig:single_coating_index}(b)
shows the corresponding influence of the
coating's refractive index, $n_1$, 
on the effective sphere's refractive index,
$n_p^\ast$.
Unlike the diameter, 
the change in refractive index,
$\Delta_n = n_p^\ast - n_0$,
does not depend 
linearly on $n_1$.
Generally speaking, a low-index coating
with $n_1 < n_0$ reduces the effective refractive
index, $n_p^\ast < n_0$, while a high-index
coating increases it.
The overall magnitude of this effect is
less than one part per thousand for the conditions
considered, which suggests that trends
in the effective refractive index 
are not likely to provide a practical
basis for detecting and
characterizing individual molecular overlayers.

\subsection{Fractional coverage and partial coatings}
\label{sec:coverage}

Binding sites may not cover the surface
of a probe bead uniformly, nor need
target molecules fill all of the available
binding sites.
Such incomplete coverage is depicted schematically in Fig.~\ref{fig:geometry}(a).
If target molecules with refractive index
$n_1$ fill a fraction, 
$f$, of the available sites,
the remainder of the surface layer is filled
with the fluid medium at refractive index
$n_m$.
The surface layer then has an
effective refractive index, $n_1^\ast$,
intermediate between
$n_1$ and $n_m$ that is 
accounted for by Maxwell Garnett effective-medium
theory \cite{markel_introduction_2016}:
\begin{subequations}
\label{eq:fillingfactor}
\begin{equation}
    L(n_1^\ast/n_m) = f L(n_1/n_m),
\end{equation}
where
\begin{equation}
    L(m) = \frac{m^2 - 1}{m^2 + 2}
\end{equation}
\end{subequations}
is the Lorentz-Lorenz function.
From this, we obtain an expression for
the effective refractive index of the
partial coating
\begin{align}
    \frac{n_1^\ast(f)}{n_m}
    & =
    \sqrt{
    \frac{1 + 2 f L(n_1/n_m)}{1 - f L(n_1/n_m)}} \\
    & \approx
    1 + \frac{3}{2} f L(n_1/n_m)
    \quad\text{for $f L(n_1/n_m) < 1$.}
\end{align}

The assumed linear dependence of
$d_p^\ast$ on the refractive index of the coating
therefore suggests that the effective
diameter of a probe bead
depends on the filling factor, $f$, as
\begin{align}
    d_p^\ast 
    & = 
    d_0 + 
    2 a_1 \frac{n_1^\ast(f) - n_m}{n_0 - n_m} 
    \label{eq:exactfractional} \\
    & \approx
    d_0 +
    3 a_1 f \frac{L(n_1/n_m)}{n_0/n_m - 1} .
    \label{eq:approximatefractional}
\end{align}
We predict, therefore, that the 
measured increase in the
effective diameter is
roughly proportional to the
fractional coverage of bound molecules.

\begin{figure*}
    \centering
    \includegraphics[width=0.7\textwidth]{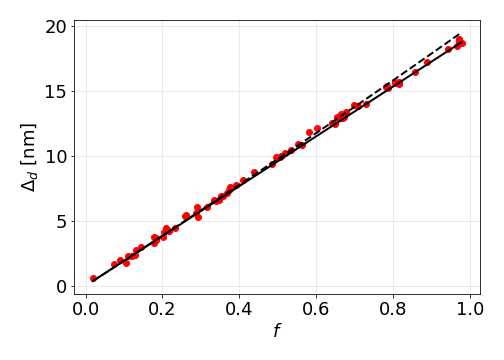}
    \caption{Effective shift in diameter from the effective-sphere model, $\Delta_d$, plotted against fraction of surface coverage, $f$ for a sphere
    of diameter $d_0 = \SI{1}{\um}$ and refractive
    index $n_0 = \num{1.60}$ coated with a \SI{10}{\nm}-thick layer with $n_1 = \num{1.60}$. 
    Coverage fraction is computed using Maxwell Garnett effective-medium theory, with $f = 0$ corresponding to $n_1^\ast = n_m$ corresponding and $f = 1$ corresponding
    to $n_1^\ast = n_1$.
    Dashed and solid curves correspond to predictions of
    Eq.~\eqref{eq:exactfractional} and Eq.~\eqref{eq:approximatefractional}, respectively.}
    \label{fig:fractional_cover}
\end{figure*}

Figure~\ref{fig:fractional_cover} presents
the representative case of $n_1 = n_0 = \num{1.6}$, 
with a coating thickness of $a_1 = \SI{10}{\nm}$.
This should be a reasonable model for a polystyrene
sphere coated with a monolayer of protein.
Discrete (red) data points represent results
of numerical experiments for randomly selected
filling fractions, ranging from
a bare sphere at $f = 0$ to a complete
coating at $f = 1$.
The solid line
is a comparison with the linear approximation
from Eq.~\eqref{eq:approximatefractional}, which
interestingly agrees better with the data than
the dashed curve representing the 
full expression from Eq.~\eqref{eq:exactfractional}.

\subsection{Molecular binding assays: Double coatings}

The foregoing results show that the effective-sphere
model reasonably models the light-scattering
properties of the probe beads used for
holographic molecular binding assays.
We next address how those properties change
when target molecules occupy the binding
sites on the surface of a probe bead
to form a second layer, as 
depicted in Fig.~\ref{fig:geometry}(a).
We continue to choose $d_0 = \SI{1}{\um}$
and $n_0 = \num{1.6}$ to model the micrometer-diameter
polystyrene substrate bead used in experimental studies.
Once the coating of binding sites is added, the probe
beads have effective diameter $d_0^\ast$ and
effective refractive index $n_0^\ast$, both of
which are determined by fitting to the
effective-sphere model.
Adding a layer of target molecules on top of this
constitutes a model for a binding assay with
effective properties $d_p^\ast$ and $n_p^\ast$.
The standard assay involves monitoring the
difference, $\Delta d_p^\ast = d_p^\ast - d_0^\ast$,
in the probe beads' effective diameter upon
binding.
We also monitor the change in refractive index,
$\Delta n_p^\ast = n_p^\ast - n_0^\ast$.

\begin{figure*}
    \centering
    \includegraphics[width=\textwidth]{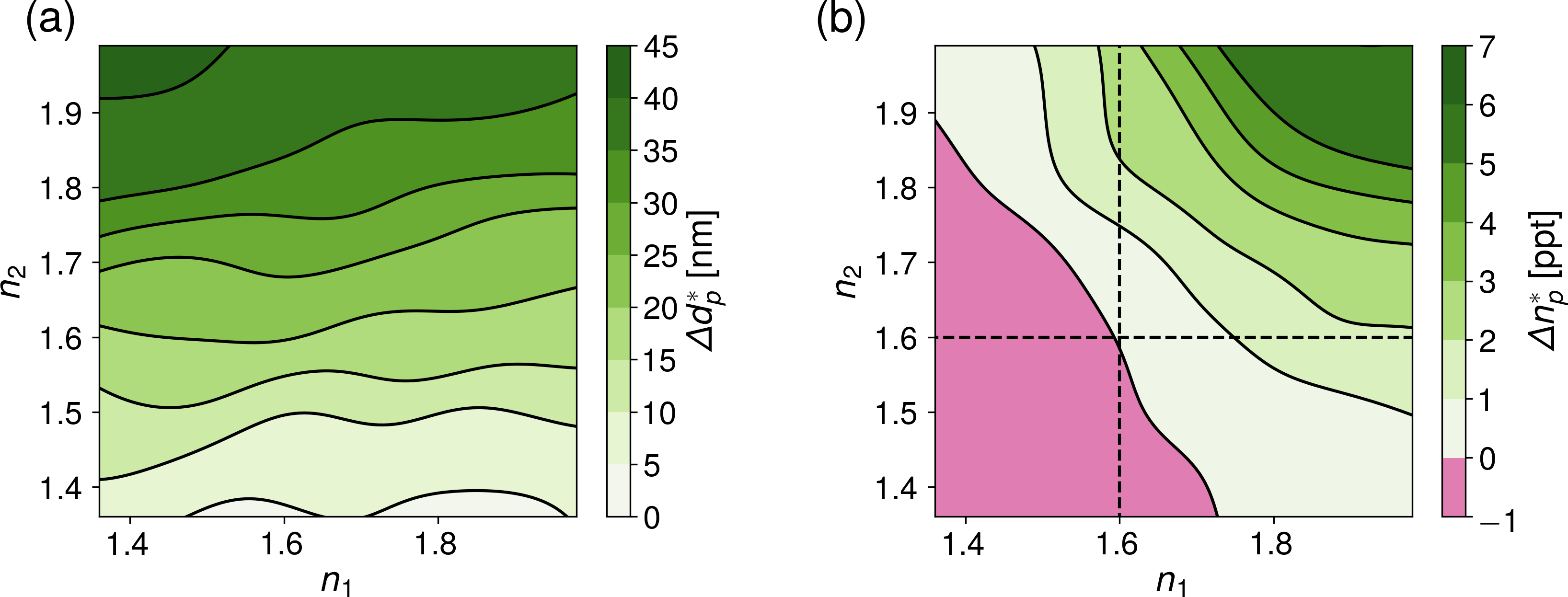}
    \caption{Differences in effective sphere parameters between a singly-coated and doubly-coated sphere, varying values of coating effective refractive indices $n_1$ and $n_2$. 
    (a) $\Delta d_p^\ast$ is shown to be largely independent of the index of the first coating, $n_1$.
    (b) $\Delta n_p^\ast$ can be either positive or negative depending on coating parameters, but is close to zero when $n_1 = n_0, n_2 = n_0$  (dashed lines), where $n_0 = \num{1.6}$. We observe a larger shift in effective index when the coating indexes are above that of the underlying sphere.}
    \label{fig:fitdifference}
\end{figure*}

For concreteness, we choose the two coatings 
to have the same thickness,
$a_1 = a_2 = \SI{10}{\nm}$, 
while $n_1$ and $n_2$ are chosen at random between 
\num{1.4} and \num{1.7}.
This range of refractive indexes matches
expectations for protein coatings 
\cite{mcmeekin_refractive_1964}
given that
the the coatings may not be complete.
Performing two fits to the effective-sphere model 
for each parameter pair $(n_1, n_2)$ yields 
measurements of $\Delta d_p^\ast$ and
$\Delta n_p^\ast$
that are presented in
Fig.~\ref{fig:fitdifference}(a) and
Fig.~\ref{fig:fitdifference}(b), respectively.

Binding-induced increases in the
effective diameter, $\Delta d_p^\ast$, 
are found to be largely independent of the 
inner coating's refractive index, $n_1$.
Changes in the apparent size depend much
more strongly on the refractive index
of the outer coating, $n_2$.
This is fortunate for practical molecular binding
assays because it means that variations in the
density of binding sites from bead to bead
will not contribute disproportionately to
variations in $d_p^\ast$ and therefore in
estimates for the filling fraction, $f$.
This, in turn, increases the reliability of
measurements of the concentration of target molecules
based on holographic measurements of $d_p^\ast$.

Binding-induced
changes in the bead's refractive index are
far more subtle than changes in the size.
The influence of the outer coating on both
the apparent size and refractive index
increases as
the refractive index of the coating becomes more mismatched with
the refractive index of the substrate bead.
These observations
suggests that the sensitivity of
molecular binding assays
can be improved by reducing the
refractive index of the substrate beads.
Previous experiments have used commercial
polystyrene substrate beads with 
relatively high refractive indexes,
$n_0 = \num{1.60}$.
Better choices
might include silica
with a refractive index around
$n_0 = \num{1.42}$, 
poly(methyl methacrylate) (PMMA) with a refractive
index around \num{1.50}, and
3-(trimethoxysilyl)propyl methacrylate (TPM)
with a refractive index around
$n_0 = \num{1.51}$ \cite{van_der_wel_preparation_2017,middleton_optimizing_2019}.

\section{Comparison with Experiments}
\label{sec:experiments}

\begin{figure*}
    \centering
    \includegraphics[width=\textwidth]{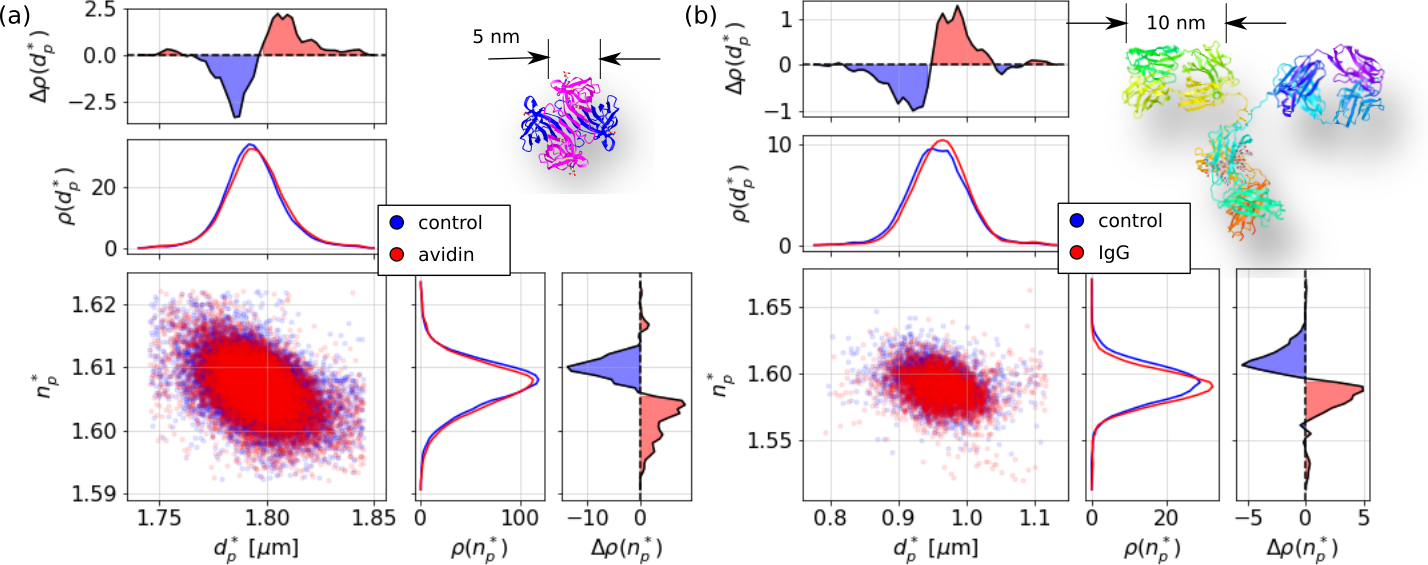}
    \caption{Holographic particle characterization data
    for (a) avidin binding to biotinylated polystyrene spheres and
    (b) IgG binding to polystyrene
    spheres coated with protein A.
    Data from Ref.~\cite{zagzag_holographic_2020}.
    The scatter plots show the distributions
    of single-particle characterization results.
    Control data from the probe beads are
    colored blue. Results after binding are colored
    red.
    Each data set also shows the projected
    distributions of particle diameters,
    $\rho(d_p)$, and refractive indexes,
    $\rho(n_p)$, as well as the differences
    $\Delta \rho(d_p)$ and $\Delta \rho(n_p)$
    in measured diameter and refractive index, respectively,
    caused by molecular binding.
    Both cases show an increase in effective
    diameter and a decrease in effective
    refractive index after binding.
    Insets show structures of
    (a) tetrameric avidin (PDB code: 2AVI) and
    (b) IgG (PDB code: 1IGT).}
    \label{fig:IgG}
\end{figure*}

Insights gained from these numerical studies
are useful for interpreting experimental
realizations of holographic molecular
binding assays \cite{cheong_flow_2009,zagzag_holographic_2020}.
The data in Fig.~\ref{fig:IgG} are reproduced from
Ref.~\cite{zagzag_holographic_2020}
and show how molecular binding changes
effective-sphere properties
in two cases: avidin binding to biotinylated 
polystyrene spheres (Fig.~\ref{fig:IgG}(a)) and IgG 
binding to polystyrene spheres coated with protein A 
(Fig.~\ref{fig:IgG}(b)). 
Each point in the scatter plot represents
the measured diameter and
refractive index of a single colloidal particle
dispersed in a buffer containing dissolved target molecules.
The dark (blue) data points represent the
properties of the probe beads, $d_0^\ast$ and $n_0^\ast$,
before incubation. The light (red) points represent
properties of the same population of beads
after incubation, $d_p^\ast$ and $n_p^\ast$.
To facilitate comparisons, we also plot
projected probability distributions, 
$\rho(d_p^\ast)$ and $\rho(n_p^\ast)$,
of the measured diameters and refractive indexes,
respectively, both before and after incubation.
The differences in these distributions,
$\Delta \rho(d_p^\ast) 
= \rho(d_p^\ast) - \rho(d_0^\ast)$ and
$\Delta \rho(n_p^\ast)
= \rho(n_p^\ast) - \rho(n_0^\ast)$,
emphasize shifts in the distributions
due to incubation with target molecules.

The biotinylated probe spheres 
have a mean diameter of 
$d_0^\ast = \SI{1.7935(4)}{\um}$.
This increases to
$d_p^\ast = \SI{1.7956(5)}{\um}$ after
tetrameric avidin binds to the beads.
Confidence intervals for these values represent
the uncertainty in the mean of more than 
\num{15000} particles in each data set
and therefore are much smaller than the
uncertainty in a single-bead measurement.
The observed shift of 
$\Delta d_p^\ast = \SI{1.1(1)}{\nm}$ 
is smaller than the \SI{5}{\nm}
domain size of avidin.
Whereas the beads' diameter increases
upon binding, their measured refractive
indexes decrease from
$n_0^\ast = \num{1.60730(3)}$ to
$n_p^\ast = \num{1.60693(3)}$,
a net change of
$\Delta n_p^\ast = \SI{-0.37(4)}{ppt}$.
The probe beads' refractive index is consistent
with expectations for polystyrene, presumably
because biotin is a small molecule; a coating
of biotin therefore does not substantially affect
the substrate beads' light-scattering
properties.

The protein A-coated spheres have a mean diameter of 
$d_0^\ast = \SI{0.9573(8)}{\um}$ before incubation 
with the antibody IgG.
This increases to $d_p^\ast = \SI{0.9622(6)}{\um}$ after 
\SI{45}{\minute} incubation resulting in a shift of
$\Delta d_p^\ast = \SI{4.9(10)}{\nm}$.
This increase is larger than was observed
for the avidin-biotin system presumably
because IgG is substantially larger
than tetrameric avidin.
Once again, however, $\Delta d_p^\ast$
is much smaller
than the size of the target molecule.
Uncertainties are larger in this case
because the statistical ensemble consists
of only \SI{3000}{particles} per sample.

As for the biotin-avidin system,
binding IgG causes the probe beads' refractive index to
shift downward 
from $n_0^\ast = \num{1.5926(3)}$
before binding to
$n_p^\ast = \num{1.5897(2)}$ after,
a decrease of
$\Delta n_p^\ast = \SI{-2.9(3)}{ppt}$.
The initial refractive index of the
protein A coated probe beads is smaller
than expectations for polystyrene
presumably because of the influence
of the protein. Protein A is nearly as
large as tetrameric avidin and might be
expected to have a comparably sizable
influence.

In both cases, binding with target
molecules leads to an increase in 
the holographically measured particle
diameter that is smaller than the physical size
of the target molecules and a decrease
in the measured refractive index.
Looking to the results in Fig.~\ref{fig:fitdifference},
these trends can be explained if the
coatings of bound molecules have lower
refractive indexes than the effective
refractive index of the substrate beads.
Specifically, we interpret these results
to show that both avidin and IgG have
refractive indexes substantially smaller
than \num{1.6} at the densities of the
experimentally obtained coatings.
Choosing substrate beads with lower refractive
indexes therefore should increase the
apparent shift in bead diameter upon
binding thereby increasing the target molecules'
influence on holographically measurable
properties and improving the sensitivity
of the assay.

\section{Conclusions}

We have used the discrete-dipole approximation 
to model label-free bead-based
molecular binding assays performed with
holographic particle characterization in
the effective-sphere approximation.
Our computational study confirms that 
interpreting the holograms of coated sphere
with the Lorenz-Mie theory for homogeneous spheres
yields valuable information on the presence
and characteristics of the coatings
while retaining the demonstrated speed
and robustness of standard holographic
particle characterization.
Our study validates previous experimental
reports of holographic molecular binding
assays \cite{cheong_flow_2009,zagzag_holographic_2020}
and explains trends in those measurements
as arising from the mismatch in refractive
index between the substrate beads and the
molecular coatings.
Because this mismatch depends on the filling
factor, $f$, of bound molecules on the
beads' surfaces, the particles' effective
diameters and refractive indexes change
continuously as binding proceeds.
We have shown that changes in the diameter
scale linearly with $f$ to a very good
approximation. This means that trends in the
holographically measured diameter can be
mapped directly onto trends in the 
fraction of filled binding sites.

Our results furthermore provide guidance for 
optimizing holographic molecular binding assays.
Most notably, the sensitivity of such assays
to variations in the filling factor, $f$, of
the available binding sites can be increased
by reducing the refractive index of the 
substrate beads.
The polystyrene substrates used for many standard
bead-based assays are not the best choice
for this application, therefore, because their
refractive index is quite high.
Alternatives such as silica, PMMA and TPM spheres
offer attractive alternatives for
such assays.
Fluorinated latex spheres might be
an exceptionally good choice 
\cite{sacanna_microemulsion_2004}.
The choice of substrate for holographic
molecular binding assays therefore can be
optimized both for optical properties
and also for physical properties such
as buoyancy to facilitate processing
of tests.
Beads of different sizes and compositions
can be functionalized with different
binding sites and combined into a multiplexed
assay. The individual tests can be monitored
in parallel through the unique ability
of holographic particle characterization
to differentiate particles by both size
and refractive index.
These considerations should be particularly
useful for designing serological and
diagnostic tests for viral infection,
with immediate urgency begin placed on
addressing the ongoing COVID-19 pandemic.

The open-source software used for this study is available
at
\url{https://github.com/laltman2/DDA_binding_assays}.

\section*{Funding}

National Science Foundation (NSF) (1420073, 2027013). 

\section*{Acknowledgments}

This work was supported primarily
by the MRSEC program of
the National Science Foundation through 
Award No.\ DMR-1420073.
Additional funding was provided
by the RAPID program of the National Science Foundation
through Award No.\ DMR-2027013.

\section*{Disclosures}

DGG: Spheryx, Inc. (F,I,P,S).



\end{document}